\documentstyle[epsf]{article}

\newcommand{\beq}{ \begin{equation} }
\newcommand{\eeq}{ \end{equation} }
\newcommand{\bea}{ \begin{eqnarray} }
\newcommand{\eea}{ \end{eqnarray} }
\newcommand{\be}{ \beta }
\newcommand{\f}{ \frac }
\newcommand{\de}{ \partial }

\begin{document}
\thispagestyle{empty}
\parskip=12pt
\raggedbottom

\def\mytoday#1{{ } \ifcase\month \or
 January\or February\or March\or April\or May\or June\or
 July\or August\or September\or October\or November\or December\fi
%\space\number\day ,
 \space \number\year}
\noindent
\hspace*{9cm} BUTP--96/13\\
\vspace*{1cm}
\begin{center}

{\LARGE SU(3) Thermodynamics on Small Lattices}

\vspace{1cm}

Alessandro Papa \\
Institute for Theoretical Physics \\
University of Bern \\
Sidlerstrasse 5, CH--3012 Bern, Switzerland

\vspace{0.5cm}

\mytoday \\ \vspace*{0.5cm}

\nopagebreak[4]

\begin{abstract}
The free energy density of the SU(3) gauge theory at temperatures 
$T/T_c = 4/3, 3/2$ and 2 is calculated on lattices with temporal extent as
small as $N_\tau = 2, 3$ and spatial extent $N_\sigma = 4 N_\tau$ 
using parametrized fixed point actions. Although cut--off effects are seen, 
they are hugely suppressed with respect to Wilson and Symanzik--improved 
actions and at $N_\tau=3$ there is already a good agreement with the 
continuum limit as extrapolated from the results with the Wilson action
at $N_\tau = 6 $ and 8.
\end{abstract}

\end{center}

\vspace{0.5cm}

PACS number(s): 05.70.Ce, 11.10.Wx, 11.10.Hi, 12.38.Gc, 12.38.Mh

\eject
 
\section{\bf Introduction}
\label{sec:intro}

Understanding the high temperature phase of QCD is of great importance
especially in astronomy and cosmology. In such a phase
quarks and gluons are expected to behave like a plasma of weakly
interacting particles approximately described by an ideal gas.
To find signatures of this plasma is the central goal of present and 
forthcoming experiments in heavy--ion collisions at CERN and BNL. 

Owing to the well--known infrared problem~\cite{Lin80} of finite temperature
QCD, the perturbative expansions of thermodynamic quantities can be 
calculated only up to $g^5$ order and their convergence is very bad even at
temperatures much larger than $T_c$~\cite{AZ94,ZK95}. Lattice
numerical simulations are therefore an essential tool in investigating
QCD at high temperatures, where non--perturbative effects seem to be
important at least in the vicinity of the phase transition. 

In lattice numerical simulations, the temperature $T$ and 
the volume $V$ are determined by the lattice size $N_\sigma^3\times
N_\tau$, $N_\tau < N_\sigma$ through  
\beq
T \; = \; \f{1}{N_\tau a} \;\;\; , \;\;\;\;\;\; 
V \; = \; (N_\sigma a)^3 \;\;\; ,
\label{eq:T,V}
\eeq
and (anti--)periodic boundary conditions have to be imposed in the time 
direction for (Fermi) Bose fields.
The lattice spacing $a$ depends on the coupling $\be=2N/g^2$; the asymptotic 
form of this dependence can be known analytically from perturbation theory.

At high temperatures, high momentum modes give relevant contributions to
the free energy density. In this regime, thermodynamic 
quantities are therefore strongly influenced by the finite lattice cut--off 
$a^{-1}$ introduced by the discretization of the theory. This cut--off 
effect manifests itself in $O(a^2)$ (i.e. $ O(1/N_\tau^2)$ at fixed $T$) 
corrections which vanish in the continuum limit. 
How large $N_\tau$ has to be in order to allow reliable extrapolations to 
the continuum depends on the specific regularization of the theory, i.e. on 
the lattice action.

The Wilson plaquette action~\cite{Wil74} is the simplest action which can be 
used in numerical simulations of the pure gauge theory. 
It has however the disappointing drawback of
introducing large cut--off dependences in physical observables:
in zero temperature simulations, the continuum limit is reached
at lattice spacings smaller than 0.1 fm, so physics at the hadronic 
scales can be simulated only on uncomfortably large lattices.
One can have a feeling of the relevance of cut--off effects with the
SU(N) Wilson action looking at the expressions for the energy 
density $\epsilon$ and the pressure $p$ of the 
gas of gluons in the $T\rightarrow\infty$ limit~\cite{EKR95,BKL95}
\beq
\f{\epsilon}{T^4} = \f{3p}{T^4} = (N^2-1) \f{\pi^2}{15}
\left[ 1 + \f{30}{63}\left(\f{\pi}{N_\tau}\right)^2 +
\f{1}{3}\left(\f{\pi}{N_\tau}\right)^4 + O\left(\f{1}{N_\tau^6}\right)
\right] \;\;\;.
\eeq
The deviation from the Stefan--Boltzmann law $(N^2-1)\pi^2/15$ is very
prominent: for $N_\tau = 4$, where also $O(1/N_\tau^6)$ contributions are 
not negligible, the deviation from the continuum is about 50\%~\cite{BKL95}.
The relevance of cut--off effects in SU(3) thermodynamic quantities 
at temperatures $T$ equal to a few times $T_c$ is less 
dramatic~\cite{BEKLLLP95,Kar95}, but large enough
to call for lattices up to $32^3\times 8$ before safely extrapolating to the 
continuum. Since fluctuations in thermodynamic quantities go like
$1/\sqrt{V_4} \sim N_\tau^{-2}$, a constant accuracy requires
a computational effort which increases at least like $N_\tau^{10}$ (here,
a fixed $N_\sigma/N_\tau$ ratio has been assumed).

Adopting the Symanzik program~\cite{Sym83} to improve actions
reduces indeed lattice artifacts. Since the improvement procedure is
perturbative, the benefit is expected only at large values of the coupling 
$\be$, i.e. at large temperatures $T$ for fixed $N_\tau$. Indeed, in SU(N)
lattice gauge theory in the $T\rightarrow\infty$ limit, the first correction 
to the Stefan--Boltzmann law can be made $O(1/N_\tau^4)$, if the action is 
improved at the tree level with a $1\times 2$ or a $2 \times 2$ plaquette, 
or even $O(1/N_\tau^6)$, if it is improved with a
$2 \times 2$ and a $3 \times 3$ plaquette~\cite{BKL95}. For $N_\tau =4$
the deviation from the Stefan--Boltzmann law is $1.34\%$ with the $1\times 2$
improvement and $8.77\%$ with the $2\times 2$ one~\cite{BKL95}.

As a radical solution of the problem of cut--off effects in lattice 
simulations, P.~Hasenfratz and F.~Niedermayer~\cite{HN94} have recently 
suggested to use perfect lattice actions, i.e. actions which are 
completely free of cut--off effects. The existence of such 
actions is well established in the Wilson's renormalization group (RG)
theory~\cite{Wil74a}. The fixed point (FP) action and the renormalized 
trajectory of a RG transformation define a perfect quantum action. For an 
asymptotically free theory, the FP action itself reproduces the properties 
of the continuum classical action -- it is the perfect classical action.
A formal argument of perturbative RG implies that FP actions are also 
1--loop (quantum) perfect~\cite{Wil80,DHHN95a}. FP actions are therefore the 
first step in the ambitious program of building perfect lattice actions.

In Ref.~\cite{HN94} a general method for the determination of FP actions for
asymptotically free theories was proposed, which was successfully applied 
to the 2--d O(3) non--linear $\sigma$--model on the lattice. 
A parametrization of the FP action suitable for numerical simulations was 
found showing no cut--off effects up to correlation lengths as small as 
$\xi \sim 3$ lattice spacings, i.e. well away from the continuum limit. 
Moreover, a perturbative calculation~\cite{FHNP95} of the mass gap in a 
periodic box of size $L$, $m(L)$, using the lattice FP action showed that 
cut--off effects are absent also on the 1--loop level.

A similar program has been carried out for SU(3) lattice gauge
theory~\cite{DHHN95a,DHHN95b} starting from two different RG transformations
(type I and type II in the following). A few--parameter approximation
has also been suggested~\cite{DHHN95b} for both type I and type II FP actions
in terms of the plaquette and the twisted perimeter--six loop.
Preliminary scaling tests of the type I FP action 
on the torelon mass and the static $q\bar{q}$ potential, using the critical
deconfining temperature to set the scale, have already shown a 
clear improvement with respect to the Wilson action~\cite{DHHN95b}.

The aim of this paper is to perform a study of the cut--off
effects of pa\-ra\-me\-tri\-zed FP actions in bulk thermodynamic 
quantities of the SU(3) gauge theory at finite temperature. Indeed, 
the contribution of high excitations to these quantities is important, making
lattice determinations quite sensitive to the cut--off. FP actions are 
implemented on lattices as small as $8^3\times 2$ and $12^3\times 3$ to 
calculate the free energy density at temperatures $T/T_c = 4/3, \ 3/2, \ 2$. 

The paper is organized as follows: \newline
in Section~2 the construction of a (parametrized) FP action is shortly
recalled and some comments are presented about the expected cut--off
effects in the free energy density; \newline
in Section~3 elements and techniques of lattice finite temperature SU(3) 
thermodynamics relevant to the remaining part of the paper 
are briefly reviewed; \newline
in Section~4 Monte Carlo (MC) determinations of the free energy density
with parametrized FP actions are compared with previous results with
Wilson and Symanzik--improved actions.

\section{\bf Parametrized FP actions and cut--off effects in the free energy 
density}

\begin{figure}[htbp]
\setlength{\unitlength}{.9mm}
\begin{center}
\begin{picture}(120,80)
\put(30,30){\vector(1,0){80}}
\put(30,30){\vector(0,1){50}}
\put(30,30){\vector(-1,-1){30}}
\put(4,0){$c_1$}
\put(32,76){$c_2,\ldots$}
\put(105,25){$1/\beta$}
\put(5,43){FP}
\put(15,45){\circle*{1}}
\put(15,30){\vector(0,1){10}}
\put(15,60){\vector(0,-1){10}}
\put(6,36){\vector(1,1){6}}
\put(24,54){\vector(-1,-1){6}}
\put(22,38){\vector(-1,1){4}}
\put(8,52){\vector(1,-1){4}}
\multiput(15,45)(.5,0){40}{\circle*{1}}
\multiput(35,45)(.5,.1){10}{\circle*{1}}
\multiput(40,46)(.5,.2){10}{\circle*{1}}
\multiput(45,48)(.5,.3){10}{\circle*{1}}
\multiput(50,51)(.5,.2){10}{\circle*{1}}
\multiput(55,53)(.5,.1){10}{\circle*{1}}
\multiput(60,54)(.5,0.05){60}{\circle*{1}}
\thicklines
\multiput(40,45)(10,0){5}{\line(1,0){5}}
\thinlines
\put(90,45){FP action}
\put(90,60){RT}
\multiput(50,51)(-.25,0){15}{\circle*{.5}}
\multiput(50,51)(-.13,-.21){15}{\circle*{.5}}
\multiput(70,55)(-.22,0.11){15}{\circle*{.5}}
\multiput(70,55)(-.20,-.14){15}{\circle*{.5}}
\end{picture}
\end{center}
\caption[]{Schematic flow diagram for asymptotically free gauge theories.}
\label{fig:RTflow}
\end{figure}
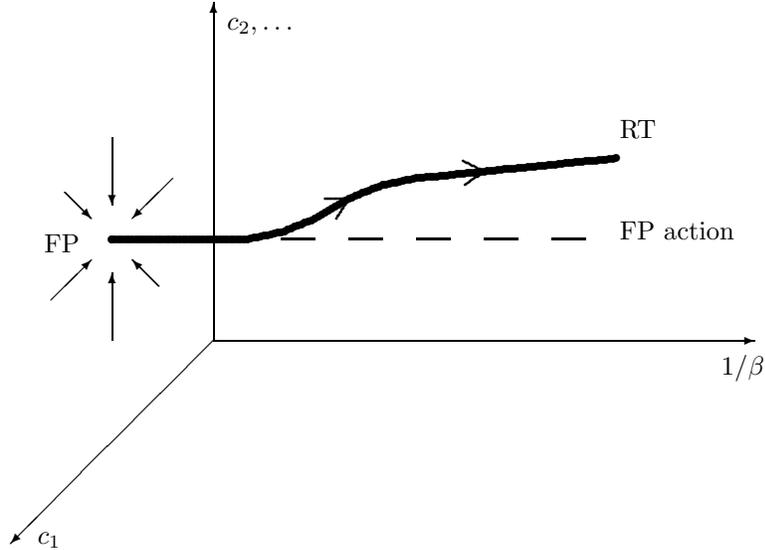

The partition function of an SU(N) gauge theory defined on a hypercubic
lattice is
\beq
Z = \int DU \: e^{-\beta A(U)} \;\;\; ,
\label{eq:part}
\eeq
where $\be A(U)$ is some lattice regularization of the continuum action
expressed in terms of products of link variables $U_\mu(n)=e^{iA_\mu(n)}
\in$SU(N) along arbitrary closed loops. If $\be$, $c_2, c_3, \ldots$ denote
the couplings of $\be A(U)$, the action can be represented by a point 
in the infinite dimensional space of the couplings (see 
Fig.~\ref{fig:RTflow}). 

Under repeated RG transformations the action moves in the space of couplings.
A RG transformation with scale factor 2 can be defined in the following way:
\beq
e^{- \be^{\prime}A^{\prime}(V)} = \int DU\: e^{-\be(A(U)+T(U,V))}\;\;\; ;
\label{eq:RG}
\eeq
here $V$ is the coarse link living on the lattice with
spacing $a^\prime=2\, a$, related to a local average of the original
link variables $U$; $T(U,V)$ is the blocking kernel which defines this 
average, normalized in order to keep the partition function (and therefore 
the free energy) invariant under the transformation. 
%If gauge invariance is to be 
%preserved by the RG transformation, the blocking kernel must satisfy 
%\beq
%T(U^g,V^g) \; = \; T(U,V) \;\;\;,
%\eeq
%where $U^g$ and $V^g$ are gauge transformed configurations on the fine and
%coarse lattice, respectively. Beyond this restriction, 
There is a wide freedom in defining the RG average; this freedom can be 
exploited for optimization in numerical simulations. 

On the critical surface $\be=\infty$, Eq.~(\ref{eq:RG}) leads to the saddle 
point equation\footnote{$\be^\prime =\be -O(1)$ owing to asymptotic 
freedom.}:
\beq
A^\prime (V) = \min_{\{U\}} [ \, A(U)+T(U,V) \, ] \;\;\; ;
\eeq
the FP point of this transformation is therefore
\beq
A^{FP} (V) = \min_{\{U\}} [ \, A^{FP}(U)+T(U,V) \, ] \;\;\; . 
\label{eq:FP}
\eeq
An important consequence of this equation is that if the configuration 
$\{V\}$ satisfies the FP classical equations  and is a local minimum
of $A^{FP} (V)$ then the fine configuration $\{U(V)\}$ minimizing the
r.h.s. of Eq.~(\ref{eq:FP}) satisfies the FP classical equations as well
and the value of the action remains unchanged, $A^{FP} (V) = A^{FP} 
(U(V))~$\cite{HN94,DHHN95a}.
As a consequence, the FP action has exact scale invariant instanton solutions
with the same action as in the continuum theory.

Although the FP lies on the critical surface $\be = \infty$,
by the expression ``FP action'' we mean $\be A^{FP}(U)$ which corresponds to 
the line which leaves orthogonally the critical surface, i.e. $\be A^{FP}(U) 
\rightarrow \be^\prime [ A^{FP}(U) + O(1/\be^2)]$. For very large $\be$, 
the FP action stays close to the RT, i.e. it is a good approximation of the 
perfect action. This follows from a formal argument~\cite{Wil80,DHHN95a} 
of perturbative RG implying that FP actions are perfect not only in the 
classical limit $\be = \infty$, but also at 1--loop order of perturbation
theory. The range of $\be$ for which the FP action and the RT stay
close together cannot be known {\it a priori} just by looking at the RG
transformation.

The FP equation~(\ref{eq:FP}) is highly non--trivial. On smooth $V$
configurations, it can be expanded in powers of the vectors
potential and studied analytically. At the lowest order in the expansion of 
the vector fields the FP action is quadratic and can be found 
analytically~\cite{DHHN95a}: it describes massless free gluons with an exact 
relativistic spectrum for all values of the momenta, i.e. not restricted to
the first Brillouin zone. 

On configurations with large fluctuations, numerical methods have 
to be used to solve Eq.~(\ref{eq:FP}). For practical reasons, however, it is
useful to find a general parametrization for the FP action, simple enough to 
be used in Monte Carlo simulations.
In Ref.~\cite{DHHN95a} two RG transformations (type I and type II) were 
defined depending both on two parameters which were optimized in order to 
make the quadratic part of the FP action the most short ranged. The type I FP
action (which is used in the following) was parametrized with powers 
of the real part of the trace of the plaquette and the twisted 
perimeter--six loop, calling for (only) 8 parameters.

The source of cut--off effects in determinations of the free energy density 
on the lattice using parametrized FP actions as regularization is 
three--fold:

i. Although the blocking kernel $T(U,V)$ is normalized in order to keep  
the partition function invariant under the RG transformation, a
field--independent term in $A^\prime (V)$ appears in Eq.~(\ref{eq:RG}) 
as a result of integration on the fine links $U$. The amount of this constant
term can be singled out plugging the flat configuration of the coarse lattice
($V_\mu(n_B) = 1$ for every site $n_B$) into Eq.~(\ref{eq:RG}): it is easy
to see that it is given by the sum of quantum fluctuations around the flat 
configuration of the fine lattice. One usually discards this
term since it does not affect the dynamics of the coarse field. For
$N_\tau,N_\sigma$ (a few times) larger than the ``size of the action'' this
term is proportional to the lattice volume $N_\sigma^3 N_\tau$ and 
gives again a constant (independent of $N_\tau$ and $N_\sigma$) contribution
to the free energy density which can be discarded as well. However, for very
small $N_\tau$ (and/or $N_\sigma$) the field--independent part will depend
non--trivially on $N_\tau$. This produces an artifact which decreases
exponentially with $N_\tau$. In practice, it will show up only for $N_\tau
=2$.  \newline
It should be stressed that all the above observations hold even in the  
classical limit. 

ii. The perfect action at finite $\be$ is represented by the points on the 
RT. At moderate values of $\be$, the FP action deviates from the RT: this
deviation causes cut--off effects. 

iii. The parametrization of a FP action introduces additional cut--off 
effects. This is especially true for the few--coupling parametrizations
needed to perform numerical simulations. In most cases parametrizations are 
reliable only in the region of correlation lengths in which the numerical 
procedure to fix the parameters was performed.

While cut--off effects referred in ii. are not fully under control,
those referred in iii. can be considerably reduced if a ``good'' 
parametrization is found. Short range actions are easier to parametrize 
than extended actions. Experience with RG suggests that a good averaging
procedure in the block transformation will decrease the extension of the
corresponding FP action. In Ref.~\cite{BN96} 
a new RG transformation has been proposed with a smoother and more rotational
symmetric smearing kernel $T(U,V)$, with a larger number of free parameters
used to optimize the resulting FP action. Preliminary scaling test of this 
type III FP action with the static $q\bar{q}$ potential have already shown an
improvement of the performances with respect to the type I FP action (using
the same 8--coupling parametrization)~\cite{BN96}.

\section{\bf SU(3) thermodynamics on the lattice}

All the continuum thermodynamic quantities for a system with zero chemical
potential can be derived from the free energy density\footnote{The Boltzmann
constant is set to one.}
\beq
f \; = \; - \f{T}{V} \ln Z(T,V) \;\;\; , 
\eeq
where $V$ is the volume of the system, $T$ the temperature and $Z$
the canonical partition function. The energy density $\epsilon$ and the 
pressure $p$ are determined by the formulae
\bea
\epsilon & = & \f{T^2}{V} \f{\de \ln Z(T,V)}{\de T} \;\;\;, \\
p & = & T \f{\de \ln Z(T,V)}{\de V} \;\;\; .
\eea
For homogeneous, large ($N_\sigma \gg \xi$) systems $\de \ln Z(T,V)/\de V =
\ln Z(T,V)/V$, so the 
pressure and the free energy density are connected by the simple relation
\beq
p = -f \;\;\; .
\label{eq:p-f}
\eeq

On the lattice, the temperature $T$ and the volume $V$ are determined
by the lattice size $N_\sigma^3 \times N_\tau$ through Eqs.~(\ref{eq:T,V}).
The partition function $Z_L$ is given by
\beq
Z_L(T,V) \; = \; \int [DU] e^{-\be A(U;N_\tau,N_\sigma)} \;\;\; ,
\eeq
where $U$ are the link variables and $\be A(U;N_\sigma,N_\tau)$ is the 
lattice action.

The free energy density can hardly be determined on the lattice directly by 
a Monte Carlo (which allows to calculate easily expectation values). It can be 
extracted however through the expectation value of the action density
$S_T = A/(N_\sigma^3 N_\tau)$, which is related simply to the derivative of 
the free energy density with respect to the bare coupling 
$\be$~\cite{EFKMW90}.
The free energy density at a given temperature $T$ can be thus determined,
apart from an additive constant, by an integration in $\be$ up to the
value corresponding to that temperature $T$ 
\beq
\left.\f{f}{T^4}\right|^\be_{\be_0} \; = \; - N_\tau^4 \int_{\be_0}^\be
d\be^{\prime} \: (\: \langle S_0 \rangle - \langle S_T \rangle \: ) \;\;\; ;
\label{eq:f}
\eeq
here the contribution from the action density $S_0$ at zero temperature has 
been subtracted in order to normalize to zero the free energy density at zero
temperature. Actually, $\langle S_0 \rangle$ is calculated on the symmetric 
lattice $N_\sigma^4$, i.e. at a small non-zero temperature.   
Since the low temperature confined phase of SU(N) gauge theory is dominated 
by glueballs which are very massive ($m_G \sim $1 GeV), the free energy 
density is expected to drop with the exponential low $e^{-m_G/T}$. This 
suggests that choosing a small enough $\be_0$ fixes the additive integration 
constant to zero (in actual numerical calculations the signal for $ N_\tau^4
(\langle S_0 \rangle - \langle S_T \rangle)$ is zero already at $\be$ 
slightly smaller than the critical coupling for the deconfining transition). 
Using Eq.~(\ref{eq:f}), the continuous curve for the free energy density 
as a function of the coupling $\be$ can be determined for a fixed 
$N_\tau$ over an arbitrary range. The delicate task is to assign a physical 
temperature $T$ to each value of $\be$ through the relation 
$T^{-1} = N_\tau a(\be)$. 
According to the specific lattice action adopted, $(T N_\tau)^{-1}$ may not 
be small enough in the temperature range of interest to allow the use of the 
2--loop asymptotic expression for $a(\be)$. In this case it is necessary to 
improve the determination of $a(\be)$ non--perturbatively, studying the 
$\be$--dependence of a physical observable\footnote{It should be checked, 
however, that $\be$ is large enough to prevent any dependence on the choice 
of the observable, i.e. the non--universality of the $\be$--function.}.

In Refs.~\cite{BEKLLLP95} numerical simulations were performed using 
the Wilson action with $N_\tau$ = 4, 6, and 8 and $N_\sigma/N_\tau \geq 4$.
Assuming that $N_\sigma/N_\tau$ was large enough to satisfy 
Eq.~(\ref{eq:p-f}), the curve showing the dependence of $p/T^4$ on the 
physical temperature $T$ was determined up to temperatures as large as 
$\sim 5 T_c$. The non--perturbative $\be$--function was obtained imposing 
the independence of the critical temperature $T_c$ on the critical couplings
for $N_\tau$ = 3,4,6,8 and 12 and using the ansatz $a \Lambda_L =
f(\be)|_{2-loop} \cdot \lambda (\be)$, where $\lambda(\be)$ is a smooth
function satisfying $\lambda (\be\rightarrow \infty) = 1$. 
It was checked that such a non--perturbative
$\be$--function is in agreement with a MCRG analysis of ratios of Wilson
loops~\cite{QCDTARO93} for $\be$ larger than 6.0, thus suggesting this
value of $\be$ as the onset of universality. According to the critical
couplings $\be_c(N_\tau)$ measured in that paper with the Wilson action,
only for $N_\tau=8$ the curve $p/T^4$ for $T/T_c \geq 1$ is completely
in the universal region of the $\be$--function. This is not true 
for $N_\tau=4,6$ at small $T/T_c$. 

In Ref.~\cite{Bei96} a similar program has been carried out using the
$2\times 2$ Symanzik--improved action on a $24^3\times 4$ lattice. The
physical scale has been set through measurements of the string tension
on the $24^4$ lattice and of the critical coupling on the $24^3\times 4$
lattice. Interestingly enough, the improvement towards the continuum limit
occurs even at temperatures slightly above $T_c$.

\begin{figure}[htb]
\begin{center}
%\vskip 10mm
\leavevmode
\epsfxsize=120mm
\epsfbox{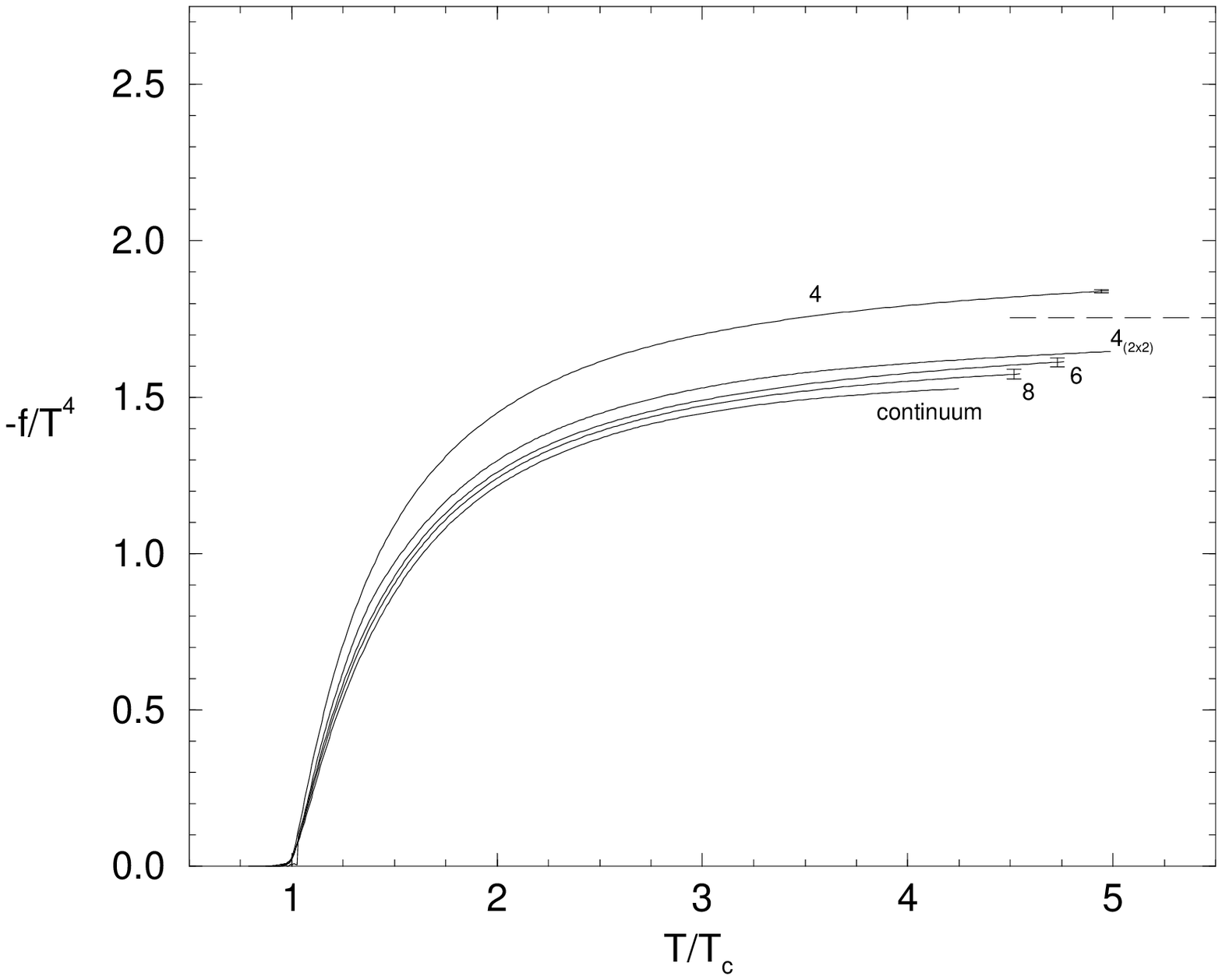}
%\vskip 10mm
\end{center}
\caption[]{The free energy density $f$ versus $T/T_c$ on lattices 
$16^3\times 4$, $32^3\times 6$ and $32^3\times 8$ with the Wilson 
action~\cite{BEKLLLP95} (curves labeled with 4, 6 and 8) and on lattice
$24^3\times 4$ with the $2\times 2$ Symanzik--improved action~\cite{Bei96} 
(curve labeled with $4_{(2\times 2)}$). The lower curve is
the extrapolation to $N_\tau \rightarrow \infty$ obtained from the curves 
with $N_\tau=6$ and 8. The horizontal dashed line indicates the ideal gas
continuum value.}
\label{fig:intro}
\end{figure}

In Fig.~\ref{fig:intro}, the present situation for lattice determinations of
the free energy density in finite temperature SU(3) gauge theory is 
summarized\footnote{I am indebted to B.~Beinlich, G.~Boyd and E.~Laermann 
for giving me tables of the results in Ref.~\cite{BEKLLLP95} and
Refs.~\cite{Bei96} necessary to build the figure.}.

\section{Monte Carlo results}

In this Section, the free energy density of finite temperature SU(3)
gauge theory is determined at temperatures $T/T_c = 4/3, 3/2$ and 2 
on lattices with $N_\tau=2,3$ and $N_\sigma =4 N_\tau$, adopting
the parametrized FP actions of two
different RG transformations. The first one is the type I RG transformation
defined in Ref.~\cite{DHHN95a}; the second one is the new 
proposal~\cite{BN96} -- type III RG transformation. Both actions are 
parametrized in terms of powers of the real part of the trace of the 
plaquette and of the twisted perimeter--six loop (i.e the loop defined by the
path $\hat{x}\hat{y}\hat{z}$-$\hat{x}$-$\hat{y}$-$\hat{z}$, for 
example). Given the following form for the parametrization
\beq
A(U) = \f{1}{N}\sum_{C, \ i\geq1} c_i(C)[N-{\rm ReTr}(U_C)]^i \;\;\; ,
\label{eq:par}
\eeq
where the sum is over all the closed paths which define a plaquette or a
twisted loop and $U_C = \Pi_C U_\mu(n)$, the couplings $c_i(C)$, 
$i=1,\ldots, 4$, are determined in order to approximate the FP action
on a large set of configurations (for details, see 
Refs.~\cite{DHHN95b,BN96}). They are summarized in Table~1.

\begin{table}[htb]

\begin{center}
Table 1: Couplings of the parametrized FP action (\ref{eq:par}) for the
RG transformations of type I~\cite{DHHN95b} and type III~\cite{BN96}.
\end{center}

\centering

\begin{tabular}{cccccc} 
\cline{1 - 6}
            &           &   $c_1$   &   $c_2$   &   $c_3$    &   $c_4$   \\
\cline{1 - 6} 
type I      & plaquette & 0.523     &  0.0021   &  0.0053    & 0 .0167   \\
            & twisted   & 0.0597    &  0.0054   &  0.0051    &--0.0006   \\
\cline{1 - 6}  
type III    & plaquette & 0.4822    &  0.2288   &--0.1248    & 0.0228    \\
            & twisted   & 0.0647    &--0.0224   &  0.0030    & 0.0035    \\
\cline{1 - 6}
\end{tabular}
\end{table}

According to Eq.~(\ref{eq:f}), the lattice determination of the free energy
density at a temperature $T$ calls first for the calculation of expectation 
values of the action density on lattices $N_\sigma^4$ and  $N_\sigma^3
\times N_\tau$ for a sufficiently large set of $\be$'s and then for an
integration up to the value of $\be$ corresponding to the temperature
$T$ at a fixed $N_\tau$. 

The usual procedure to determine the
correspondence between the coupling $\be$ and the temperature $T$ at a fixed
$N_\tau$ needs a non--perturbative determination of $a(\be)$ or,
equivalently, of the $\be$--function. This procedure usually involves an
interpolation of numerical data (see the previous Section). Since the 
central point of this paper is to single out lattice cut--off effects of FP 
actions, an alternative way which introduces no additional artificial 
effects has been exploited. The idea is that the critical coupling
$\be_c(N_\tau^\prime)$ on a lattice with $N_\tau^\prime$ sites in the time 
direction and infinite spatial volume corresponds to the physical temperature
$T = (N_\tau^\prime/N_\tau) T_c$ on a lattice with $N_\tau$ sites in the time
direction. Indeed, starting from the definition of the critical coupling 
$\be_c(N_\tau^\prime)$ 
\beq
T_c \; = \; \f{1}{N_\tau^\prime a(\be_c(N_\tau^\prime))} \;\;\; ,
\label{eq:Tc}
\eeq
it is easy to see that it corresponds to the temperature 
\beq
T \; = \; \f{1}{N_\tau a(\be_c(N_\tau^\prime))} \; = \;
\f{N_\tau^\prime}{N_\tau} T_c 
\label{eq:T}
\eeq
on a lattice with temporal extent $N_\tau$. The advantage of this procedure 
is that critical couplings can be determined with high precision. The 
drawback is, of course, that only discrete sets of temperatures can be 
determined. 

For the type I FP action, the critical couplings on lattices with
$N_\tau = 2, 3, 4, 6$ were already determined in Ref.~\cite{DHHN95b}; for
the type III FP action the same work has also been done~\cite{BN96}.
In Table~2, results for both actions have been summarized.

\begin{table}[htb]

\begin{center}
Table 2: Critical couplings at finite volume and extrapolated to infinite
volume for the FP actions type I and III parametrized as in Table~1.
\end{center}

\centering

\begin{tabular}{cccccc} 
\cline{1 - 6}
         & volume   & $N_\tau=2$ & $N_\tau=3$ & $N_\tau=4$ & $N_\tau=6$  \\
\cline{1 - 6}
type I   & $ 4^3$   &  2.89(1)   &            &            &             \\
         & $ 6^3$   &  2.945(15) &  3.320(15) &            &             \\
         & $ 8^3$   &  2.96(1)   &  3.34(1)   &  3.50(1)   &             \\
         & $10^3$   &  2.960(5)  &  3.34(1)   &  3.51(1)   &  3.78(2)    \\
         & $12^3$   &  2.975(5)  &  3.343(5)  &  3.520(5)  &  3.775(5)   \\
         & $15^3$   &            &            &            &  3.777(5)   \\
         & $18^3$   &            &            &            &  3.790(5)   \\
         & infinite &  2.976(5)  &  3.345(5)  &  3.520(5)  &  3.790(6)   \\
\cline{1 - 6}
type III & $ 4^3$   & 3.361(5)   &            &            &             \\
         & $ 5^3$   & 3.378(3)   &            &            &             \\
         & $ 6^3$   & 3.385(9)   &  3.568(4)  &            &             \\
         & $ 8^3$   & 3.395(3)   &            &  3.678(3)  &             \\
         & $ 9^3$   &            &  3.581(4)  &            &             \\
         & $10^3$   & 3.399(5)   &            &  3.686(3)  &  3.91(4)    \\
         & $12^3$   &            &  3.587(5)  &  3.691(5)  &  3.87(1)    \\
         & $14^3$   &            &            &  3.691(5)  &  3.882(7)   \\
         & $16^3$   &            &            &            &  3.882(8)   \\
         & infinite & 3.400(3)   &  3.588(4)  &  3.695(4)  &  3.885(13)  \\
\cline{1 - 6}
\end{tabular}
\end{table}

According to Eq.~(\ref{eq:T}) and making use of infinite volume
extrapolations in Table~2, the $\be$ couplings corresponding to temperatures
the $T/T_c = 3/2, 2$ and 3 on a lattice with $N_\tau = 2$ and those 
corresponding to temperatures $T/T_c = 4/3$ and 2 on a lattice with 
$N_\tau = 3$ can be singled out, for both type I and type III FP actions.

In this paper, simulations have been performed on lattices with $N_\tau=$ 2 
and 3 and with the ratio $N_\sigma/N_\tau$ fixed to 4 in order to rule out 
finite size effects\footnote{It is known from measurements of the energy
density in SU(2)~\cite{EKR95} that thermodynamic observables depend on the
physical volume through $V^{1/3}T\equiv N_\sigma/N_\tau$.}. 
The simulation algorithm adopted was a mixture of four
sweeps of a 20--hit Metropolis and one sweep of an over--relaxation 
consisting in four updates of SU(2) subgroups. 
With the same algorithm, the Wilson action would take $\sim 7$ times 
less for a sweep~\cite{DHHN95b}. 

\begin{figure}[htb]
\begin{center}
%\vskip 10mm
\leavevmode
\epsfxsize=120mm
\epsfbox{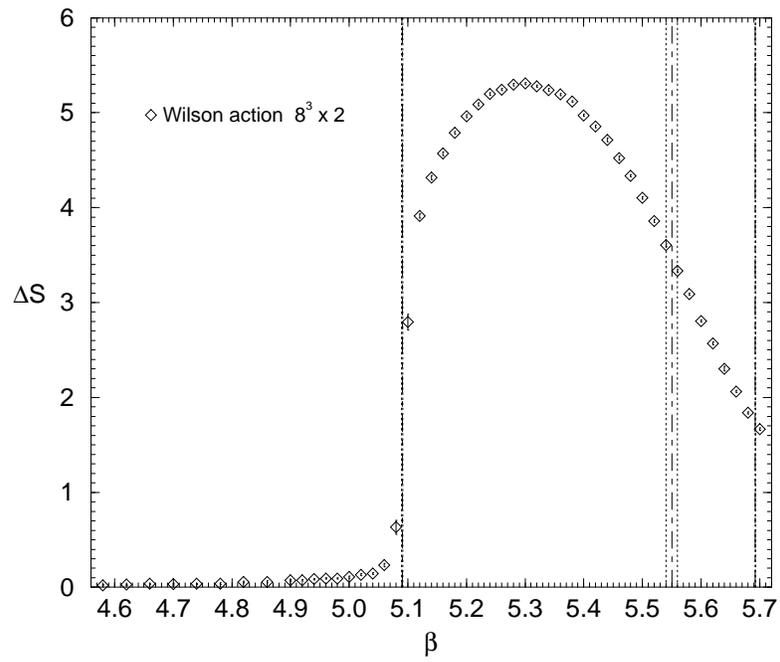}
%\vskip 10mm
\end{center}
\caption[]{$\Delta S \equiv N_\tau^4(\langle S_0\rangle -\langle S_T\rangle)$
versus $\be$ for the Wilson action
on a lattice $8^3\times 2$ lattice. The vertical lines represent the critical
couplings and the related errors on lattices $8^3\times 2$~\cite{Pen84}, 
$12^3\times 3$~\cite{FHK93} and $\infty^3\times 4$~\cite{BEKLLLP95},
respectively from left to right.}
\label{fig:wilson}
\end{figure}

\begin{figure}[htb]
\begin{center}
%\vskip 10mm
\leavevmode
\epsfxsize=120mm
\epsfbox{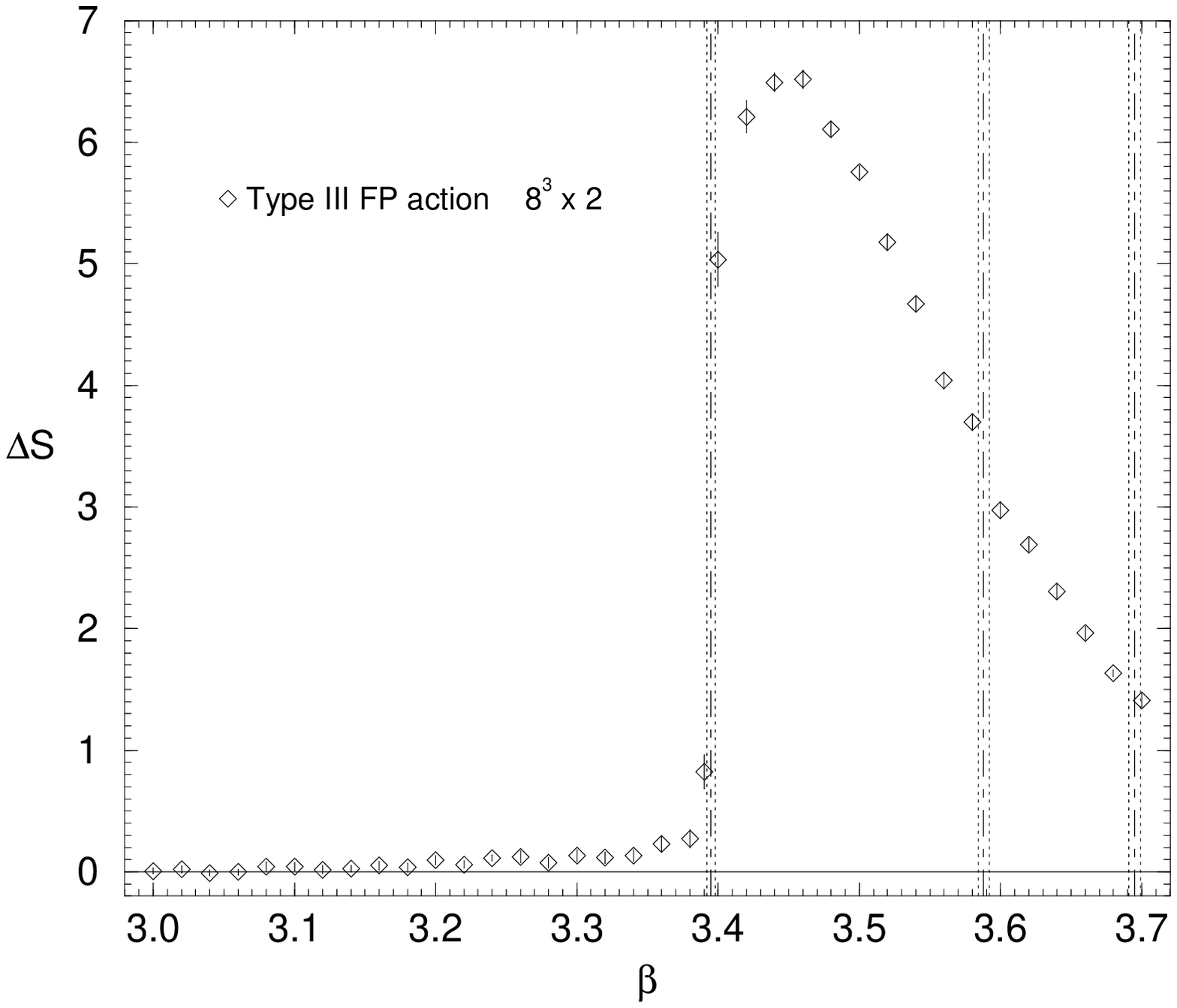}
%\vskip 10mm
\end{center}
\caption[]{$\Delta S \equiv N_\tau^4(\langle S_0\rangle -\langle S_T\rangle)$
versus $\be$ for the type III FP action on a $8^3\times 2$ lattice. The 
vertical lines represent the critical couplings and the related errors 
on lattices $8^3\times 2$, $\infty^3\times 3$ and $\infty^3\times 4$,
respectively from left to right (see Table~2).}
\label{fig:type3_2}
\end{figure}

\begin{figure}[htb]
\begin{center}
%\vskip 10mm
\leavevmode
\epsfxsize=120mm
\epsfbox{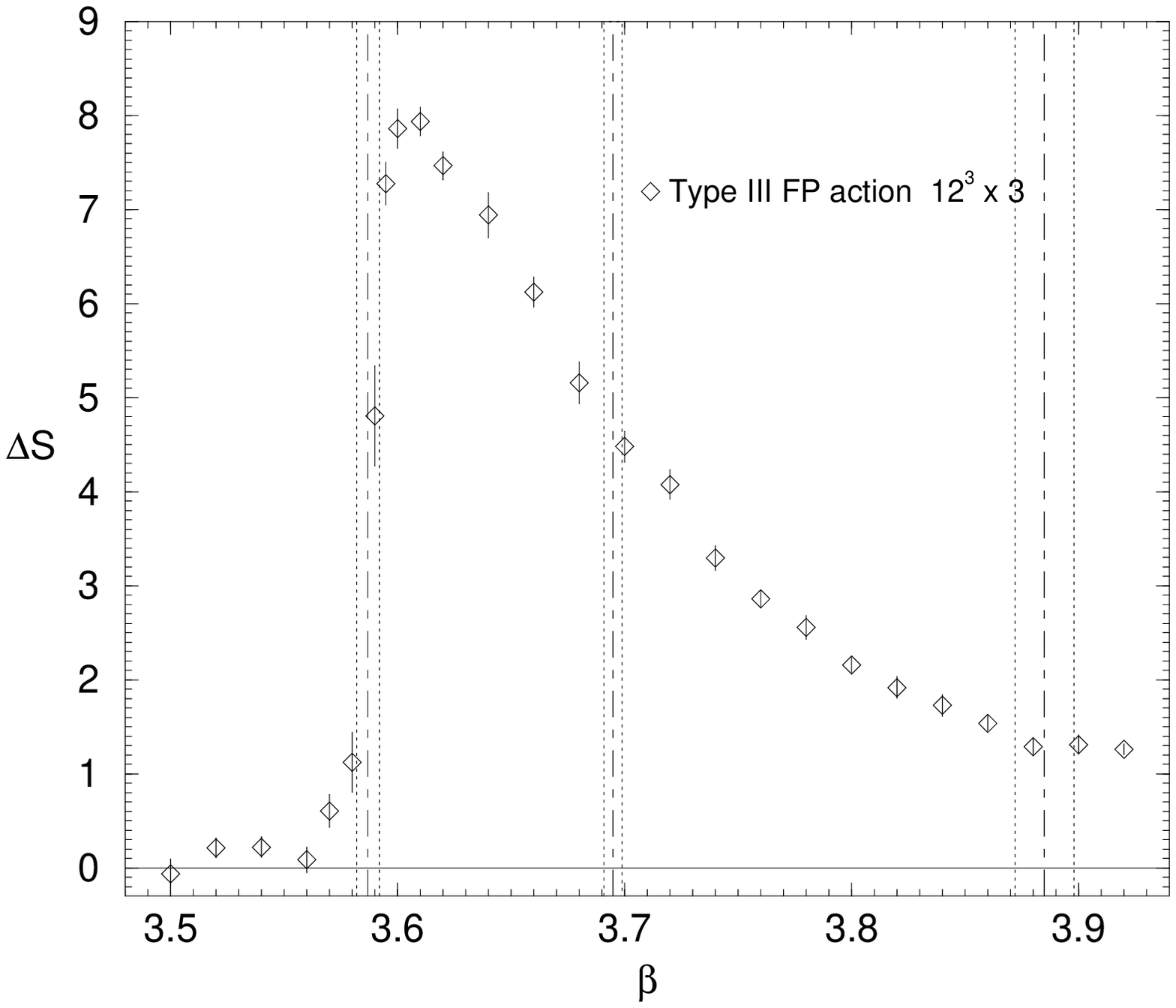}
%\vskip 10mm
\end{center}
\caption[]{$\Delta S \equiv N_\tau^4(\langle S_0\rangle -\langle S_T\rangle)$
versus $\be$ for the type III FP action on a $12^3\times 3$ lattice. 
The vertical lines represent the critical couplings and the related errors 
on lattices $12^3\times 3$, $\infty^3\times 4$
and $\infty^3 \times 6$, respectively from left to right (see Table~2).}
\label{fig:type3_3}
\end{figure}

The free energy density was determined according to 
Eq.~(\ref{eq:f}) at temperatures $T/T_c = 3/2 $ and 2 on the lattice with 
$N_\tau=2$ and at temperatures $T/T_c = 4/3$ and 2 on the lattice with 
$N_\tau=3$. This allows to compare the results at $T/T_c = 2$ obtained on 
lattices with different $N_\tau$ and to check the amount of cut--off effects.
In Figs.~\ref{fig:wilson}, \ref{fig:type3_2} and \ref{fig:type3_3}, the 
quantity $\Delta S = N_\tau^4 (\langle S_0\rangle - 
\langle S_T\rangle) $ for a large range of $\be$ is plotted for the
Wilson action with $N_\tau=2$ and for the type III FP action with
$N_\tau = 2$ and 3, respectively. In the case of the type I FP action the 
qualitative behavior is the same as for the type III FP action.
In all three cases the action density shows a rapid variation at the critical
coupling, which is smoother on the smaller lattices. The time history of 
the runs for $\langle S_T\rangle$ in the transition region clearly shows 
flips between two phases, as expected for a first order phase transition. 

For the calculation of the free energy density, $\Delta S$ has to be 
integrated with respect to $\be$. For this purpose, a spline fit
interpolation of the data for $\Delta S$ was first performed. As an 
estimate of the uncertainty in the integration coming from the interpolation 
procedure, a straight line interpolation was also considered. 
Although careful checks have been done in the time histories of the runs 
for $S_T$ in the transition region, the possibility of a (small) systematic 
error induced by hysteresis effects cannot be excluded.
In Table~3 all the results for $-f/T^4$ found in this work are summarized. 

\begin{table}[htb]

\begin{center}
Table 3: Values of $f/T^4$ at physical temperatures $T/T_c=4/3, 3/2, 2$ 
on lattices with $N_\tau=2 $ and 3 for the type I and III FP actions
and with $N_\tau=2 $  for the Wilson action.
\end{center}

\centering

\begin{tabular}{ccccc}
\cline{1 - 5}
          &   lattice    & $T/T_c=4/3$ & $T/T_c=3/2$ & $T/T_c=2$    \\
\cline{1 - 5}
type I    & $8^3\times2$ &              &  1.2226(6)  &  1.528(4)   \\
          & $12^3\times3$&  0.769(10)   &             &  1.350(10)  \\ 
\cline{1 - 5}
type III  & $8^3\times2$ &              &  1.0805(23) &  1.3279(24) \\
          & $12^3\times3$&  0.733(11)   &             &  1.221(10)) \\
\cline{1 - 5}
Wilson    & $8^3\times2$ &              &  2.2007(8)  &  2.5652(7)  \\
\cline{1 - 5}
\end{tabular}
\end{table} 

\begin{figure}[htb]
\begin{center}
%\vskip 10mm
\leavevmode
\epsfxsize=120mm
\epsfbox{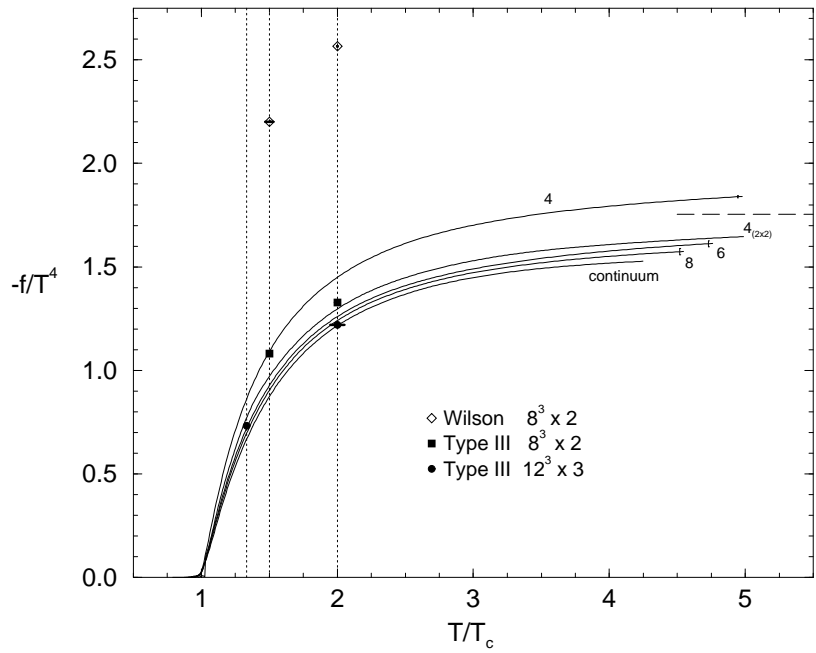}
%\vskip 10mm
\end{center}
\caption[]{The same as in Fig.~\ref{fig:intro} with the addition of the 
results of this work for the Wilson action and the type III FP action (see
Table~3). The vertical dotted lines corresponding to $T/T_c =
4/3, 3/2 $  and 2 are drawn to guide the eyes. }
\label{fig:summary}
\end{figure}

In Fig.~\ref{fig:summary} the results for the type III FP action are compared
with previous determinations with the Wilson action~\cite{BEKLLLP95} 
and the Symanzik--improved action~\cite{Bei96}, together with the continuum 
limit extrapolation~\cite{BEKLLLP95}. The horizontal error bars represent
the uncertainty in the determinations of the physical temperature $T/T_c$ 
coming from the uncertainty in the measurements of the critical couplings.
They have been estimated imposing the independence of the critical 
temperature $T_c$ on the critical couplings for the various $N_\tau$'s 
and using the ansatz $a \Lambda_L = f(\be)|_{2-loop} \cdot \lambda (\be)$, 
where $\lambda(\be)$ is a smooth function satisfying $\lambda (\be\rightarrow
\infty) = 1$. \newline
At $N_\tau=2$ there are results for $-f/T^4$ at $T/T_c = 3/2$ and 
$T/T_c = 2$:
while the Wilson action shows $O(100\%)$ cut--off effects, the type III
FP action gives values differing from the continuum values estimated
in~\cite{BEKLLLP95} by $\sim 24\%$ at 
$T/T_c=3/2$ and by $\sim 9\%$ at $T/T_c = 2$. \newline
At $N_\tau=3$, the result for $-f/T^4$ with the type III FP action at 
$T/T_c = 4/3$ differs from the continuum by $\sim 10\%$
and at $T/T_c = 2$ is consistent with the continuum value.

Although the global scenario exhibits an evident improvement with respect 
to the approach with the standard actions, it is worth to make some comments 
on the surviving cut--off effects in the results of the FP actions.
The qualitative understanding of such cut--off effects is straightforward. 
Recalling the classification at the end of Section~2, 
the effect mentioned in i. should explain the deviations between
$N_\tau =2$ and $N_\tau = 3$ results; the effect mentioned in ii. can be 
responsible of the cut--off effects at small $\be$ -- and so at small $T$ -- 
for each $N_\tau$; 
the effect mentioned in iii. can overlap to the previous one at small 
$\be$, where the parametrization of the FP action is more likely to 
fail to reproduce the true FP action. It would be interesting to evaluate 
quantitatively the contribution of the different sources of cut--off effects:
if one of them is largely dominating, it can be removed by an {\it ad hoc} 
strategy.

\section{Acknowledgements}
I thank P. Hasenfratz (who suggested this subject), M.~Blatter, G.~Boyd,  
F.~Farchioni, F.~Karsch and F.~Niedermayer for many useful discussions. 
I am indebted to P.~Hasenfratz and F.~Niedermayer for the Monte Carlo
code of the FP actions.

\end{document}